# Competitive market for multiple firms and economic crisis

Yong Tao [*]

School of Economics and Business Administration, Chongqing University, Chongqing 400044, China

**Abstract:** The origin of economic crises is a key problem for economics. We present a model of long-run competitive markets to show that the multiplicity of behaviors in an economic system, over a long time scale, emerge as statistical regularities (perfectly competitive markets obey Bose-Einstein statistics and purely monopolistic-competitive markets obey Boltzmann statistics) and that how interaction among firms influences the evolutionary of competitive markets. It has been widely accepted that perfect competition is most efficient. Our study shows that the perfectly competitive system, as an extreme case of competitive markets, is most efficient but not stable, and gives rise to economic crises as society reaches full employment. In the economic crisis revealed by our model, many firms condense (collapse) into the lowest supply level (zero supply, namely bankruptcy status), in analogy to Bose-Einstein condensation. This curious phenomenon arises because perfect competition (homogeneous competitions) equals symmetric (indistinguishable) investment direction, a fact abhorred by nature. Therefore, we urge the promotion of monopolistic competition (heterogeneous competitions) rather than perfect competition. To provide early warning of economic crises, we introduce a resolving index of investment, which approaches zero in the run-up to an economic crisis. On the other hand, our model discloses, as a profound conclusion, that the technological level for a long-run social or economic system is proportional to the freedom (disorder) of this system; in other words, technology equals the entropy of system. As an application of this new concept, we give a possible answer to the Needham question: "Why was it that despite the immense achievements of traditional China it had been in Europe and not in China that the scientific and industrial revolutions occurred?"

## 1. Introduction

Before the 20th century, almost all economists believed (dreamed) that markets operate perfectly; this viewpoint is based on the Say's law, that is, supply creates its own demand. Unfortunately, the world economic crisis of the 1930s, which led to the economic revolution of Keynes, demolished this belief. Keynesian economics enjoyed its heyday in the decades after the Second World War, but was forced out of the mainstream after failing a crucial test during the mid-seventies, when American stagflation sowed doubts about Keynesian economics. Thus, some insightful theorists attempted to resume *laissez faire* policies of relying heavily if not solely on supposedly efficient markets, the representative figures being Milton Friedman [1], Robert.

---

[*] Correspondence Email: taoyingyong2007@yahoo.com.cn

Lucas[2] etc. From that time onwards, many economists began to believe that the right way of describing economic phenomena was to integrate Keynesian economics with *laissez faire* policies supporting free markets. Unfortunately, the economic crises of East Asia in the 1990s and American sub-prime mortgage crisis in 2008 were rude reminders that the answer was not as simple. Economic crisis, like a devil, is always intertwined with the market economy. Why do the economic crises occur?

More recently, Jean-Philippe Bouchaud pointed out that[3] economics needs a scientific revolution, for example, involving physical methods. Furthermore, J. Doyne Farmer and Duncan Foley also pointed out that [4] the best economic models are of two types, namely "econometric" and "dynamic stochastic general equilibrium", both with fatal flaws. Both of them therefore promoted the idea that the economy needs agent-based modelling[5–6]. Certainly, people have realized that a realistic economic model should correspond to the natural regularities of the economic system and should further predict and guide governments to avert economic crises. In fact, many physical and mathematical methods now play significant roles in economics, so much so that new subjects such as econophysics[7–8] and complex network[9–10] have emerged. Despite these developments and the new insights, there is still no consensus on the origin of crises. Here we need to point out that, in the past, physicists have failed to capture the essence of economic ideas (for example, the central properties of perfect competition and monopolistic competition), while economists did not pay attention to modern ideas in physics (for example, spontaneously symmetry breaking). As a result, important ways of describing competitive markets have been missed by physicists and economists.

In this work, our object is to present a model for competitive markets resulting in early warnings of economic crises. To develop this model, we need to first sum up the central properties of competitive markets by taking advantages of microeconomics. The relevant mathematical techniques and physical notions originate from statistical physics and the concept of spontaneous symmetry breaking which has its origin in quantum field theory. Most importantly, in this paper we only need to focus on a long-run market, which can be, accurately, dealt with by our forthcoming methods. More detailed description of economics will be published elsewhere.

## 2. Competitive market and competitive equilibrium

Long-run competitive markets require that firms run on extremely short time intervals ("firm-run") compared to the market ("market-run"). Roughly, we can easily understand that a competitive market is composed of a vast number of interacting firms (agents) and that these firms (agents) run in many industries. If each industry can be related to a submarket, the competitive market, in general, is also composed of a number of diverse industries. In this market, how agents make decisions looks like how firms interact with one another. All these interactions make up "the economy". In the spirit of microeconomics, each firm only gains zero economic profit [11–12] in the long run; that is, the return to the firm only pays for its cost (both fixed cost and variable cost). Most importantly, supply creates its own demand. Since each firm always gains zero economic profit, these firms shall be able to freely move from one industry to another in any period without worrying about the barrier of the industry. If we consider the short-term market, these expressions (namely, zero economic profit and vanishing barrier of industry) are, obviously, not right; however, these properties hold in the long-term market. In fact, a real free market, in the long run (observed time long enough), approaches a competitive market. Therefore, a **competitive market** is a system exhibiting the following three properties [12]:

(a). A competitive market is composed of a vast number of firms and diverse industries;

(b). Each firm always gains zero economic profit and hence can freely move from one industry to another;

(c). Supply creates its own demand.

Property (c) is just the Say's law, which guarantees that a satisfactory economic model can be developed using only the supply of firms, without making any reference to demand. That is to say, any supply can be acceptable without worrying about possible excess supply (except economic crises, sees section 4). We suppose there are $n$ supply levels, that is, $\varepsilon_1 < \varepsilon_2 < ... < \varepsilon_n$. We suppose $a_i$ is the number of firms each of which provides the supply quantity $\varepsilon_i$; furthermore, we suppose these $a_i$ firms are distributed among $g_i$ industries. Thus, the total number $N$ and total supply $Y$ of firms can be given respectively by

$$N = \sum_{i=1}^{n} a_i, \qquad (1)$$

$$Y = \sum_{i=1}^{n} a_i \varepsilon_i. \qquad (2)$$

Property (b) implies that the long-run behavior of competitive markets can be regarded as a stochastic process. To understand this point, we need to provide two explanations. First, we observe the free movement of firms from one industry to another can be understood as a random walk, which is a Markov process [1]. Second, the property that each firm always gains zero economic profit implies that the supply quantity of each firm is a random variable; that is because, no matter what quantity of supply each firm provides, it always gains zero economic profit. In other words, for each firm there is no difference of providing any supply (except zero supply, sees section 4).

These two points indicate that the state in which a firm occupies one possible supply level and one possible industry is a stochastic process. That means, if we, in one period, see a firm in one supply level and one industry, we, in the next period, may see it in another supply level and another industry. Therefore, a competitive market, in essence, is a complex system. The notion that a market of economic interaction is a complex system can be traced at least as far back as Adam Smith in the late 1700s. In this system, agents, in pursuit of their own profit, persistently and frequently move from one industry to another [2], though they only gain zero economic profit in the long-term competition. That means humans tend to be over-focused in the short-term and blind in the long-term[3]. Most importantly, as for this stochastic process, statistical regularities should emerge in the behavior of large populations (namely, number of agents $N$ is large enough), just as the law of ideal gases emerges from the chaotic motion of individual molecules. To arrive at this idea, we need to give the definitions of macrostate and microstate of an economic system. These two states correspond to macro-economy and micro-economy respectively [(3)].

---

[1] The notion that firms entry and exit industries as a Markov process is not a new one and has been discussed [13–16] for many years.

[2] Agent tries his best to engage in as many industries as possible so as to guarantee profit maximization.

The **macrostate** of an economic system is specified by the number of firms in each of the supply levels $\{a_i\}$ of the economic system. Clearly, number $\{a_i\}$ completely determines the aggregate variables of the macro-economy such as $N$ and $Y$ in equations (1) and (2).

The **microstate** of an economic system is specified by the number of firms in each industry and each supply level of the economic system.

A microstate gives the most detailed description for an economic system, but we are not interested in this point. We will be interested in the number of microstates for each macrostate, $\Omega(a_i)$. That is because there will be many, many different microstates corresponding to a given macrostate: If we suppose all possible microstates of an economic system were equally probable [(4)], the real macrostate should correspond to the one which has the most microstates (namely, the most probable). In the spirit of statistical physics [17], we can easily understand that the competitive market would reach equilibrium as long as $\Omega(a_i)$ reaches an extreme value. Therefore, the definition of competitive equilibrium is given as follow.

**Competitive equilibrium** is reached if and only if $\Omega(a_i)$, which is subject to the constraints (1) and (2), satisfies:

(d). The first variation $\delta\Omega(a_i) = 0$;

(e). The second variation $\delta^2\Omega(a_i) < 0$.

Clearly, if there exists some institutional restriction (for example, planned economy, etc.), firms shall not be able to freely move from one industry to another arbitrary industry. Hence, $\Omega(a_i)$ can be regarded as a measure of freedom; that is to say, the larger $\Omega(a_i)$, the more freedom. On the one hand, consistent with the spirit of statistical physics, the solution of competitive equilibrium (d) and (e) gives the distribution among supply levels $\{a_i\}$ for the

---

[(3)] From a microstate of economic system, we may understand what quantity of supply each firm provides and which industry it is engaging in. But, from a macrostate of economic system, we only may understand the number of firms in each of the supply levels.

[(4)] That means every agent has fair chance to engage in various economic activities. It can be regarded as an axiom of the free economic system.

equilibrium state of system. On the other hand, based on the definition of microstate, computing the number of microstates $\Omega(a_i)$ is equivalent to seeking how many ways these $N$ firms can be arranged among all possible industries and supply levels. Hence, if we want to determine the number of firms $a_i$ with supply quantity $\varepsilon_i$ for all $n$ supply levels of the market, we need to count the number of ways, $\Omega(a_i)$, of arranging $N$ firms among all possible industries and supply levels. Next, we seek the numbers of microstates of perfectly competitive market and purely monopolistic-competitive market, $\Omega(a_i)_{per}$ and $\Omega(a_i)_{mon}$ respectively.

### 2.1. Perfectly competitive market

First, we focus on perfectly competitive markets. According to microeconomics, a perfectly competitive market, as an important property, requires that firms are identical and indistinguishable [12] (namely, that they produce homogeneous products), except satisfying properties (a)-(c). Homogeneous products, in microeconomics, generally hold in one industry. However, in the long run, if an agent exits an industry, then it can enter an arbitrary industry [(5)] in which there should not be differentiated products; otherwise there exists monopoly. Therefore, homogeneous products, in the long run, hold in all industries; this case can be understood as products without brands.

Clearly, to compute the number of microstates of a perfectly competitive market $\Omega(a_i)_{per}$, we may consider the arrangement of firms in each of the supply levels. For the $i$th supply level there will be $g_i$ industries containing a total of $a_i$ firms. It is convenient to depict the arrangement of $a_i$ firms among the $g_i$ industries in analogy with counting the ways of permuting $a_i$ balls among $g_i$ boxes. If we use balls and boxes to represent firms and industries respectively, the run process for a perfectly competitive market can be shown by Fig. 1.

Since these firms, in a perfectly competitive market, are indistinguishable, this issue of arrangement is identical to Bose-Einstein statistics [17,19] (sees Appendix D). Therefore, the number of ways of arranging $a_i$ firms among $g_i$ industries is $\dfrac{(a_i + g_i - 1)!}{a_i!(g_i - 1)!}$. Furthermore, the total number of ways of arranging $N$ indistinguishable firms among all possible industries and supply levels is

---

[(5)] For a perfectly competitive market, the processes that the homogeneous firms entry and exit the industries can be replaced by the birth and death processes of these firms, which obey Bose-Einstein statistics. In fact, the equivalent issue (replacement of industry and firm by city and inhabitant respectively) has been discussed [18].

$$\Omega(a_i)_{per} = \prod_{i=1}^{n} \frac{(a_i + g_i - 1)!}{a_i!(g_i - 1)!}. \qquad (3)$$

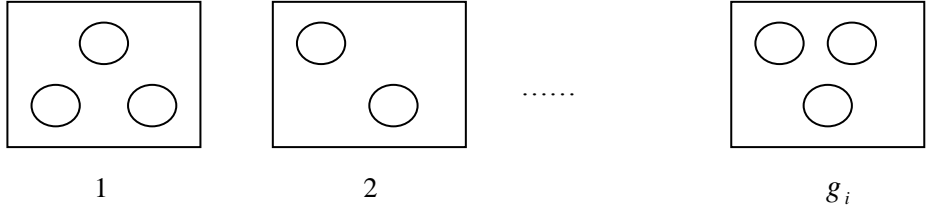

**Figure. 1 | $a_i$ indistinguishable firms are distributed among $g_i$ industries.** Here ball denotes firm and box denotes industry: Box 1 has three balls, that equals industry 1 has three firms, and so forth. $a_i$ firms move among these $g_i$ industries, that equals $a_i$ balls are permuted among these $g_i$ boxes. Since $a_i$ firms are indistinguishable, these $a_i$ balls shall not be marked by any serial number.

## 2.2. Purely monopolistic-competitive market

Second, we focus on purely monopolistic-competitive markets. A purely monopolistic-competitive market, which is different from perfectly competitive market except for properties (a)-(c), requires that firms are completely distinguishable[12] (namely, differentiated products, which can be understood by diverse brands). Likewise, we can use the Fig. 2 to show this process.

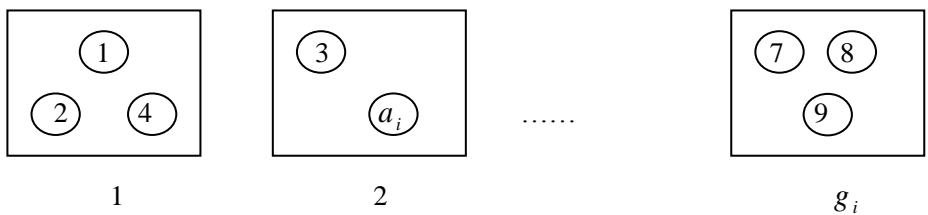

**Figure. 2 | $a_i$ distinguishable firms are distributed among $g_i$ industries.** This figure is similar to the Fig. 1 except that these $a_i$ balls are marked by serial number, which equals that these $a_i$ firms are completely distinguishable.

Since these firms, in a purely monopolistic-competitive market, are completely distinguishable, this issue of arrangement is identical to Boltzmann statistics[19]. Therefore, the total number of ways of arranging $N$ distinguishable firms among all possible industries and supply levels is

$$\Omega(a_i)_{mon} = \frac{N!}{\prod_{i=1}^{n} a_i!} \prod_{i=1}^{n} g_i^{a_i} . \qquad (4)$$

## 3. Firm's distribution of competitive market

With the constraints (1) and (2) into equations (3) and (4), the method of Lagrange multiplier gives [19]

$$a_i(I) = \frac{g_i}{e^{\alpha+\beta\varepsilon_i} - I} \begin{cases} I = 0 & (purely\ monopolistic\ competition) \\ I = 1 & (perfect\ competition) \end{cases}, \qquad (5)$$

where, $\alpha$ and $\beta$ are Langrange multipliers.

Equation (5) gives, for every supply level of the market $\varepsilon_i$, the number of firms which are distributed among $g_i$ industries for an equilibrium state. It is easy to check that equation (5) satisfies the definitions of competitive equilibrium, i.e., (d) and (e). To investigate the exact expressions of $\alpha$ and $\beta$, we need to study the aggregate supply function $Y = Y(N,T)$, where we have supposed that the aggregate supply function has only two endogenous variables, that is, total number of firms $N$ and the social technology level $T$. Technology level determines social productivity and the number of firms determines the amount of labor and capital [6]; that is to say, product is determined by labor, capital and technology.

Complete differential of $Y(N,T)$ gives

$$dY(N,T) = \mu dN + \theta dT, \qquad (6)$$

where $\mu$ (namely $\frac{\partial Y}{\partial N}$) denotes the marginal labor-capital return and $\theta$ (namely $\frac{\partial Y}{\partial T}$) the marginal technology return.

Using equations (5) and (6), calculations in Appendix A give

$$\alpha = -\frac{\mu}{\lambda\theta}, \qquad (7)$$

$$\beta = \frac{1}{\lambda\theta}, \qquad (8)$$

$$T = \lambda\left(\ln W - \alpha\frac{\partial \ln W}{\partial \alpha} - \beta\frac{\partial \ln W}{\partial \beta}\right), \qquad (9)$$

---

[6] An important role of firm is collecting labor and capital [11]: a unit of firm implies a unit of labor and capital.

where $W = W(\alpha, \beta) = \prod_i (1 - Ie^{-\alpha - \beta \varepsilon_i})^{-\frac{g_i}{I}}$, and $\lambda$ is a positive constant.

Substitution of equations (7) and (8) into equation (5) gives the distribution equation of firms,

$$a_i(I) = \frac{g_i}{e^{\frac{\varepsilon_i - \mu}{\lambda \theta}} - I}. \qquad (10)$$

For $I = 1$, equation (10) describes perfect competition; for $I = 0$, equation (10) describes purely monopolistic competition. $I = 1$ determines completely indistinguishable firms and $I = 0$ determines completely distinguishable firms. From these two cases, we guess $\omega = 1 - I$ denotes the resolution ratio: $I = 1$, which gives $\omega = 0$, describes a situation where firms cannot be distinguished at all; $I = 0$, which gives $\omega = 1$, describes a situation where firms can be completely distinguished; $0 < I < 1$, which gives $0 < \omega < 1$, describes a situation where $\omega N$ firms can be distinguished and the others can not be distinguished. If this conjecture holds, $0 \leq I < 1$ shall describe monopolistic competition. In fact, if the index $I$ were a continuous variable, we can prove this conjecture. However, the index $I$ can be understood as measure of competition, that is, the higher the value of $I$, the more fierce is competition. $I = 1$ (perfect competition) implies ultimate competition (white-hot). Here, it is easy to show $\frac{\partial Y}{\partial I} \geq 0$, that is, the more fierce the competition, the better efficiency of the market. In other words, a perfectly competitive market is the most efficient.

## 4. The stability of competitive market

Next, we shall assert that, in all cases of competitive markets, only perfect competition is unstable, and further gives rise to economic crises as society reaches full employment and the number of industries decreases (In the next section, we shall explain why the number of industries decreases when $I$ approaches 1). To this end, we need to introduce two definitions, that is, rational economic agent and positive marginal technology return.

(f) Rational economic agent[12] requires $\varepsilon_i \geq \mu$, $i = 1, 2, \ldots, n$.

The economic meaning of statement (f) is that the firms (agents) enter a market if and only if they can gain the marginal labor-capital return at least created by them (to pay for the cost); otherwise they will make a loss.

(g) Positive marginal technology return requires $\theta > 0$.

The statement (g) indicates that technical progress always gives rise to increasing output.

First, we prove that perfectly competitive market is unstable.

Proof. Using equation (10), equation (1) can be written as

$$N(I) = N(I)_{\varepsilon = \mu} + N(I)_{\varepsilon > \mu}, \qquad (11)$$

where $N(I)_{\varepsilon = \mu} = a_1(I) = \frac{g_1}{e^{\frac{\varepsilon_1 - \mu}{\lambda \theta}} - I}$, $N(I)_{\varepsilon > \mu} = \sum_{i=2}^{n} \frac{g_i}{e^{\frac{\varepsilon_i - \mu}{\lambda \theta}} - I}$.

Expression (f) shows $\varepsilon_1 = \mu$; thus, $N(I)_{\varepsilon=\mu}$ has no meaning at $I=1$. To guarantee $N(I=1)_{\varepsilon=\mu}$ makes sense, we can define $\lim_{I \to 1} g_1 = 0$ (However, note that the function structure of $g_1$ can not be determined). Nevertheless, then $N(I=1)_{\varepsilon=\mu}$ may be an arbitrarily positive number, since $N(I=1)_{\varepsilon=\mu} = \frac{0}{0}$ is an indeterminate limit and the function structure of $g_1$ can not be determined. That is different from the case of monopolistic competition ($0 \leq I < 1$) where $N(I)_{\varepsilon=\mu}$ takes a definite value. This conclusion implies that the perfectly competitive market is unstable. □

Clearly, the above proof is based on the fact that $I=1$ is a singular point of equation (10); however, as for $0 \leq I < 1$, the market is stable, that is, the monopolistically competitive market is stable.

Second, we prove that perfect competition $(I=1)$ will induce economic crises as society reaches full employment and the number of industries decreases.

Before this proof, we need to introduce the following three notions, namely, diminishing marginal labor-capital return, full employment and bankruptcy.

(h). Diminishing marginal labor-capital return requires $\frac{\partial \mu}{\partial N} = \begin{cases} \frac{\partial \mu}{\partial N} \geq 0, N \leq N_C \\ \frac{\partial \mu}{\partial N} < 0, N > N_C \end{cases}$, that is, as long as the input of labor and capital exceeds a critical value $N_C$, the marginal labor-capital return shall decline. Expression (h) is a classic hypothesis of microeconomics [11–12].

(i). Full employment requires $\mu = 0$, that is, the spare capacity of society equals zero. (Then, any new input of labor and capital will not increase any output.)

(j). In the long-run competition, a firm is bankrupt if and only if its supply is zero. (This definition originates from the study as for the Arrow-Debreu model [20], sees Appendix B.)

Proof. We suppose the total number of firms $N$ has exceeded the critical value $N_C$ so that $\frac{\partial \mu}{\partial N} < 0$; sees statement (h). As $N$ increases, $\mu$ tends to zero; nevertheless, when $\mu = 0$, society arrives at full employment. Most importantly, $\mu = 0$ requires $N = N(I=1)$, then, is a constant (namely, full employment); otherwise, as $N$ increases, $\mu$ will decrease so that

$\mu < 0$ [7]. That means full employment is a saturation state.

At the point $I = 1$ and $\mu = 0$, equation (11) can thus be written as

$$N(I=1) = N(I=1)_{\varepsilon=0} + N(I=1)_{\varepsilon>0}, \qquad (12)$$

where, $N(I=1)$ is a constant, $N(I=1)_{\varepsilon=0}$ is an indeterminate limit and

$$N(I=1)_{\varepsilon>0} = \sum_{i=2}^{n} \frac{g_i}{e^{\frac{\varepsilon_i}{\lambda\theta}} - I}. \qquad (13)$$

Once the number of industries decreases (that is to say, $g_i$ decreases), then, from equations (12) and (13), $N(I=1)_{\varepsilon>0}$ will decrease and $N(I=1)_{\varepsilon=0}$ will increase.

Increasing $N(I=1)_{\varepsilon=0}$ indicates that the number of zero supply firms increases, that is, from definition (j), the number of bankrupt firms increases. In other words, economic crises occur.
□

This state of crisis in which many firms condense (collapse) into the lowest supply level (zero supply) can be regarded as a Bose-Einstein condensation[19] of the economic system. In general, the Bose-Einstein condensation, as an important phase transition, also appears in other competitive systems. For example, complex networks can undergo Bose-Einstein condensation, in which a single node captures a macroscopic fraction of links[21].

## 5. Early warning of economic crises

To guarantee that $I = 1$ (perfect competition) does induce economic crises, we, as mentioned at the beginning of the last section, need to explain why $g_i$ decreases as $I$ approaches 1. To this end, we shall give a very important notion, that is, spontaneous breaking of investment direction.

(k). If $I = 1$, spontaneous breaking of investment direction makes investors' confidence decline and the number of invested industries hence decreases.

To understand statement (k), some explanations are required. Perfect competition $(I = 1)$ requires that all firms are indistinguishable from each other; this implies which firm shall be invested makes no difference (or equivalently, which share is chosen makes no difference). Thus, in the long run, agents will be confused for the investment direction, since which share is chosen

---

[7] The state $\mu < 0$ is no meaningful, since any input of labor and capital, then, give negative output.

makes no difference. Because of this reason, investors will decrease their investment so that the number of invested industries decreases, that is, $g_i$ decreases. (Sees Appendix D)

In fact, this is a common phenomenon in the world. For example, in a featureless snowfield, one needs a color target for orientation (otherwise, they shall get snow-blindness); in nature, spontaneous breaking gives rise to mass [22–23]. Nature, in essence, abhors absolute symmetry [24]. Perfect competition means symmetric investment direction, that is because, every firm (share) is indistinguishable. In a perfectly competitive market, agents will be confused (blind), since they find which firm is chosen for investment makes no difference. Keynes deems that the panic of investors originates from the primitive instinct of humans; in contrast, we think that investor panic originates from nature abhorring absolute symmetry.

So far, we have completely confirmed that the origin of economic crises is perfect competition. Competitive markets do perfectly operate (as Adam Smith has pointed out [25]) except in the extreme case of perfect competition.

In fact, here our economic model has pointed out a way of predicting and averting economic crises. To obtain early warnings of economic crises, we need to construct a resolving index of investment. Take the example of the stock market [8], and suppose there are $H$ shares and $m$ investors, and the $j$th investor can identify $H_j$ shares. This implies that the resolution ratio of shares is denoted by

$$\omega = \frac{\frac{H_1}{H} + \frac{H_2}{H} + ... + \frac{H_m}{H}}{m} = \frac{\sum_{j=1}^{m} H_j}{mH}, \qquad (14)$$

where, we call $\omega$ the "resolving index of investment". Clearly, $\omega = 0$ gives $I = 1$ and $\omega = 1$ gives $I = 0$; moreover, whether $\omega = 1 - I$ or not, $0 < \omega \leq 1$ (namely, $\omega H$ shares can be distinguished and others not) shall describe monopolistically competitive market. The measure of $\omega$ is very important: once $\omega$ approaches 0, a crisis looms.

The process that economic system evolves along the resolving index of investment $\omega$ and finally collapses at $\omega = 0$ is shown by Fig. 3 (where we take an example of four firms). For a large number of firms, this process, perhaps, can be simulated by using agent-based modelling [6].

## 6. Technology, freedom and Needham question

Finally, we give an answer to the Needham question [26–27] by using our theory. Calculation in Appendix C gives

---

(8) In fact, the stock market (more specifically, the bull market) comes close to perfect competition. That is the reason why economic crises almost always start from collapse of stock market.

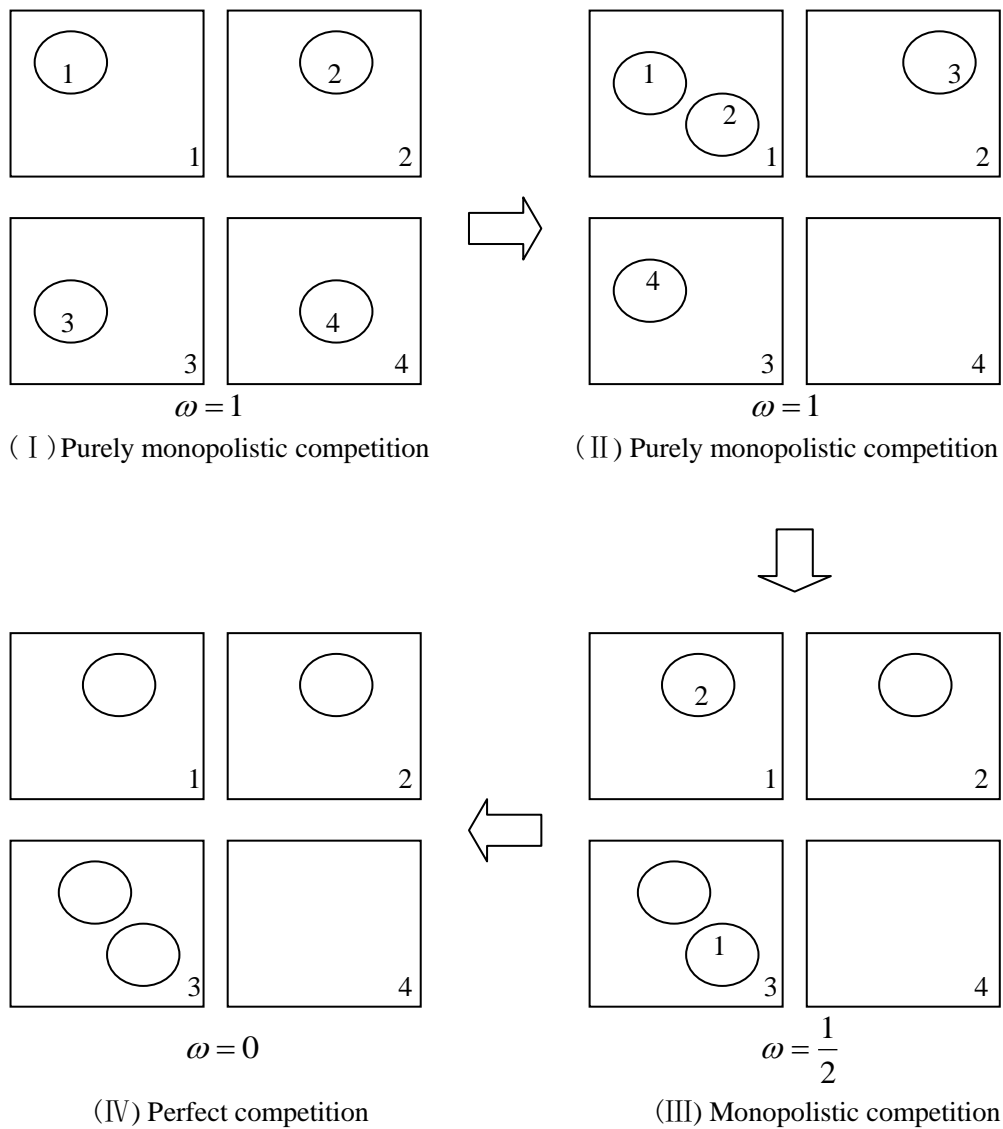

**Figure. 3 | Evolutionary process of competitive system.** Here we use the example that four firms (balls) are distributed among four industries (boxes) to show this evolutionary process. Evolutionary process of competitive market is determined by resolving index of investment $\omega$. When competitive market has evolved from (Ⅰ) to (Ⅱ), there is still $\omega = 1$ since these four firms are marked by serial number (namely, completely distinguishable). That means, not only (Ⅰ) but also (Ⅱ), describes purely monopolistic competition. When competitive market has evolved from (Ⅱ) to (Ⅲ), competition starts to be more fierce; then there are only two distinguishable firms left in (Ⅲ). The case for (Ⅲ) describes monopolistic competition, that is, $0 < \omega = \dfrac{1}{2} < 1$.

If competitive market, finally, has evolved from (Ⅲ) to (Ⅳ), then this system shall be not stable and further collapse since these four firms are completely indistinguishable. That is to say, then, there is $\omega = 0$ (perfect competition): This case corresponds to economic crisis.

$$\ln \Omega = \ln W - \alpha \frac{\partial \ln W}{\partial \alpha} - \beta \frac{\partial \ln W}{\partial \beta}. \qquad (15)$$

Comparing equations (9) and (15), we can have $T = \lambda \ln \Omega$. (16)

Equation (16) shows that, for a long-run social system, technology is a measure of freedom (disorder); that is, the more freedom, and the higher technology level. From statistical physics[17], we can realize that $T$ is just entropy.

Needham asked the provoking question: "Why was it that despite the immense achievements of traditional China it had been in Europe and not in China that the scientific and industrial revolutions occurred?"

According to statistical physics, the entropy of an isolated system shall not decrease [19], that is, the technology level for an isolated social system is almost always improving. Thus, this provides a possible explanation as to why, immense achievements were found before the 14th century not only in Europe but also in China. That is, because the entropy of an isolated system (traditional China) does not decrease. Development of technology of traditional China originates from the long-term accumulation of information (increasing entropy).

Most importantly, why did not the scientific and industrial revolutions occur in China, but in Europe? From equation (16), we have observed that technology is proportional to freedom. As for traditional China, the autocratic monarchy means there were not enough freedom for people most of the time; the few exceptions include the relatively open Tang Dynasty (the strongest period in China). However, restriction of freedom, from equation (16), retarded the development of technology of traditional China. In contrast, renaissance flourished in Europe since the 14th century, which gave rise to humanism[28] and hence broke the medieval shackles of Europe, overwhelmingly improved the freedom of European system whose technology hence made remarkable progress so that scientific and industrial revolutions finally occurred. Thus, we have provided a possible answer to the Needham question.

It is important to note that equation (16) only holds for a long-run system. Some planned economic systems, such as the Soviet Union, were able to improve their technology greatly in the short run; however, they stagnated in the long run (even system collapsed). As for free systems, such as United States and some European countries, the technology almost always are improving.

## 7. Conclusion

Diverse actions of human society, on a long time scale, will emerge as statistical regularities, just as the law of ideal gases emerges from the chaotic motion of individual molecules, A long-run competitive market, in essence, is a hugely complex system involving stochastic movements of firms: the diverse actions of different players in this system, on a long time scale, emerge as statistical regularities. Our study shows that, in all cases of competitive markets, only a perfectly competitive system is not stable and would give rise to economic crises as society reaches full employment. This fact can explain why economic crises almost always start from the collapse of the stock market, since the stock market (more specifically, the bull market) comes close to perfect competition. On the one hand, we urge that government should encourage monopolistic competition, for example, prolonging patent terms, increasing differentiated products, etc. On the other hand, we suggest a resolving index of investment for early warning of economic crises: Here we also appeal to those researchers who may be interested in our suggestion to collect relevant data to test our economic model. Another important conclusion of our economic model is that technology level for a long-run social system is proportional to freedom (disorder) of this system. This new concept may resolve the Needham question.

# Acknowledgement

The author is grateful to Professor Kenneth Young for providing many helpful comments and improving English writing. The author is grateful to the anonymous referee for providing many helpful comments. The author also thanks Mr. Wang, X. for his some suggestions with improving English writing.

# Appendix A

We construct $W(\alpha,\beta) = \prod_i \left(1 - Ie^{-\alpha-\beta\varepsilon_i}\right)^{-\frac{g_i}{I}}$.  (A.1)

Using equations (1), (2), and (5), differential of equation (A.1) gives

$$N = -\frac{\partial \ln W}{\partial \alpha},  \quad (A.2)$$

$$Y = -\frac{\partial \ln W}{\partial \beta}.  \quad (A.3)$$

Using equation (A.3), we have

$$\beta dY = \beta d\left(-\frac{\partial \ln W}{\partial \beta}\right) = -d\left(\beta \frac{\partial \ln W}{\partial \beta}\right) + d\ln W - d\left(\alpha \frac{\partial \ln W}{\partial \alpha}\right) + \alpha d\left(\frac{\partial \ln W}{\partial \alpha}\right),  \quad (A.4)$$

where, we have used formula $d \ln W = \frac{\partial \ln W}{\partial \alpha} d\alpha + \frac{\partial \ln W}{\partial \beta} d\beta$.

Substitution of equation (A.2) into equation (A.4) gives

$$dY = -\frac{\alpha}{\beta} dN + \frac{1}{\beta} d\left(\ln W - \alpha \frac{\partial \ln W}{\partial \alpha} - \beta \frac{\partial \ln W}{\partial \beta}\right).  \quad (A.5)$$

Comparing equation (A.5) and equation (6), we have

$$\alpha = -\frac{\mu}{\lambda\theta},$$

$$\beta = \frac{1}{\lambda\theta},$$

and $T = \lambda\left(\ln W - \alpha \frac{\partial \ln W}{\partial \alpha} - \beta \frac{\partial \ln W}{\partial \beta}\right),$

where, $\lambda$ is a positive constant.

## Appendix B

To give a definition for bankruptcy, we need to check the Arrow-Debreu model[20]. Here we shall confirm that the state that many firms are bankrupt is a feasible equilibrium within the framework of the Arrow-Debreu model.

Consider an Arrow-Debreu economy $\varepsilon = (X_i, \leq_i, e_i, \theta_{ij}, Y_j)$ and its concomitant economy $\hat{E} = (\hat{X}_i, \leq_i, e_i, \theta_{ij}, \hat{Y}_j)$.

Clearly, there exist $\overline{w}_i \in X_i$ and $\overline{w}_i \ll e_i$; furthermore, $(\overline{w}_1, \ldots, \overline{w}_m, 0, \ldots, 0)$ should be an attainable state, since $0 \in Y_j$. On the other hand, there had to be $\overline{w}_i \in \hat{X}_i$.

If we consider $y_j \in -Y_j^+ \cap \hat{Y}_j$ and $\sum_{i=1}^{m} \overline{w}_i \leq \sum_{i=1}^{m} e_i + \sum_{j=1}^{n} y_j$ and if we construct $M_j = -Y_j^+ \cap \hat{Y}_j$, then there will be a competitive equilibrium vector $z$ in economy $E^*$, where $E^* = (\hat{X}_i, \leq_i, e_i, \theta_{ij}, M_j)$ and $z = (\overline{x}_1, \ldots, \overline{x}_m, 0, \ldots, 0)$.

In the short run, equilibrium vector $z$ shall not have any problems; however, it is hard to believe that firms do not produce anything at all in the long run. That means, there are not any products and the limited gift shall be exhausted. If this state occurs, we can easily understand that there exists economic crisis!

Equilibrium vector $z$ reminds us that the definition of bankruptcy can be given as follow:

In the long-run competition, firm is bankrupt if and only if its supply is zero.

## Appendix C

Using the Stirling's formula, that is, $\ln m! = m(\ln m - 1), (m \gg 1)$

equations (3) and (4), except a constant factor $\ln N!$, can be written as an uniform formula, that is,

$$\ln \Omega = \sum_{i=1}^{n} \left( a_i + \frac{g_i}{I} \right) \ln(Ia_i + g_i) - \sum_{i=1}^{n} a_i \ln a_i - \sum_{i=1}^{n} \frac{g_i}{I} \ln g_i, \quad (C.1)$$

where, $I = 0$ corresponds to equation (4) and $I = 1$ corresponds to equation (3).

Substitution of equation (5) into equation (C.1) gives

$$\ln \Omega = \sum_{i=1}^{n} \left\{ -\frac{g_i}{I} \ln\left(1 - Ie^{-\alpha - \beta\varepsilon_i}\right) + a_i(\alpha + \beta\varepsilon_i) \right\}. \quad (C.2)$$

Furthermore, substitution of equation (A.1) into equation (C.2) gives

$$\ln \Omega = \ln W - \alpha \frac{\partial W}{\partial \alpha} - \beta \frac{\partial W}{\partial \beta},$$

where, we have used equations (A.2) and (A.3).

## Appendix D

**Long-run competitive market and statistical physics.** Two properties of a long-run competitive market, namely (a) and (b), imply that a competitive market is a complex system and that run of firms, namely moving among all possible industries and supply levels, can be regarded as a stochastic process. Furthermore, in the long run, we note that firms in a perfectly competitive market are identical and indistinguishable and that firms in a purely monopolistic-competitive market are completely distinguishable (Perfect competition and purely monopolistic competition are two extreme cases of competitive markets respectively). Hence, if we suppose firm corresponds to particle, supply level corresponds to energy level, and industry corresponds to quantum state, then, in the spirit of statistical physics, we can realize that these two extreme cases can be described by Bose-Einstein statistics [17,19] and Boltzmann statistics [19] respectively. Using these two statistical ways and the property (c) of competitive markets, we can construct a complete model of competitive markets.

**Spontaneously symmetry breaking.** Our model shows that, in all cases of competitive markets, only a perfectly competitive system is not stable. Furthermore, if agents, in a perfectly competitive system which reaches full employment, decrease their investment so that the number of invested industries decreases, then there will be a lot of bankrupt firms (namely economic crisis). To guarantee that the number of invested industries does decrease for perfect competition, we need to realize that perfect competition equals indistinguishable invested directions, that is, which firm shall be invested makes no difference. Indistinguishable invested directions imply symmetric invested directions. According to the idea of spontaneously symmetry breaking which has its origin in quantum field theory, agents have to break this symmetry and hence only choose some special industries or drop investment; that is to say, the number of invested industries decreases. Therefore, we conclude that perfect competition gives rise to economic crises.